\documentclass[twocolumn]{aastex63}
\usepackage[utf8]{inputenc}
\usepackage{comment}
\usepackage{amsmath}

\newcommand{\zp}{$z_\textrm{phot}$}
\newcommand\mean[1]{$\langle #1\rangle$}    

\begin{document}
\title[An Updated View of the Star-Forming Main Sequence]{Implications of a Temperature Dependent IMF II: An Updated View of the Star-Forming Main Sequence} 

\author[0000-0003-3780-6801]{Charles L. Steinhardt}
\affiliation{Cosmic Dawn Center (DAWN)}
\affiliation{Niels Bohr Institute, University of Copenhagen, Lyngbyvej 2, K\o benhavn \O~2100, Denmark}

\author[0000-0002-5460-6126]{Albert Sneppen}
\affiliation{Cosmic Dawn Center (DAWN)}
\affiliation{Niels Bohr Institute, University of Copenhagen, Lyngbyvej 2, K\o benhavn \O~2100, Denmark}

\author{Basel Mostafa}
\affiliation{California Institute of Technology, 1200 E. California Blvd., Pasadena, CA 91125, USA}
\affiliation{Cosmic Dawn Center (DAWN)}

\author{Hagan Hensley}
\affiliation{California Institute of Technology, 1200 E. California Blvd., Pasadena, CA 91125, USA}
\affiliation{Cosmic Dawn Center (DAWN)}

\author{Adam S. Jermyn}
\affiliation{CCA Flatiron}

\author{Adrian Lopez}
\affiliation{California Institute of Technology, 1200 E. California Blvd., Pasadena, CA 91125, USA}
\affiliation{Cosmic Dawn Center (DAWN)}

\author[0000-0003-1614-196X]{John Weaver}
\affiliation{Cosmic Dawn Center (DAWN)}
\affiliation{Niels Bohr Institute, University of Copenhagen, Lyngbyvej 2, K\o benhavn \O~2100, Denmark}

\author{Gabriel Brammer}
\affiliation{Cosmic Dawn Center (DAWN)}
\affiliation{Niels Bohr Institute, University of Copenhagen, Lyngbyvej 2, K\o benhavn \O~2100, Denmark}

\author[0000-0003-3873-968X]{Thomas H. Clark}
\affiliation{California Institute of Technology, 1200 E. California Blvd., Pasadena, CA 91125, USA}
\affiliation{Cosmic Dawn Center (DAWN)}

\author{Iary Davidzon}
\affiliation{Niels Bohr Institute, University of Copenhagen, Lyngbyvej 2, K\o benhavn \O~2100, Denmark}
\affiliation{Cosmic Dawn Center (DAWN)}

\author[0000-0002-6459-8772]{Andrei C. Diaconu}
\affiliation{California Institute of Technology, 1200 E. California Blvd., Pasadena, CA 91125, USA}
\affiliation{Cosmic Dawn Center (DAWN)}

\author{Bahram Mobasher}
\affiliation{University of California, Riverside, 900 University Ave. Riverside, CA 92521}

\author[0000-0001-7633-3985]{Vadim Rusakov}
\affiliation{Niels Bohr Institute, University of Copenhagen, Lyngbyvej 2, K\o benhavn \O~2100, Denmark}
\affiliation{Cosmic Dawn Center (DAWN)}

\author{Sune Toft}
\affiliation{Niels Bohr Institute, University of Copenhagen, Lyngbyvej 2, K\o benhavn \O~2100, Denmark}
\affiliation{Cosmic Dawn Center (DAWN)}

\begin{abstract}
The stellar initial mass function (IMF) is predicted to depend upon the temperature of gas in star-forming molecular clouds.  The introduction of an additional parameter, $T_{IMF}$, into photometric template fitting, allows galaxies to be fit with a range of IMFs.  Three surprising new features appear: (1) most star-forming galaxies are best fit with a bottom-lighter IMF than the Milky Way; (2) most star-forming galaxies at fixed redshift are fit with a very similar IMF; and (3) the most massive star-forming galaxies at fixed redshift instead exhibit a less bottom-light IMF, similar to that measured in quiescent galaxies.  Additionally, since stellar masses and star formation rates both depend on the IMF, these results slightly modify the resulting relationship, while yielding similar qualitative characteristics to previous studies. 
\end{abstract}





\section{Introduction}
\label{sec:intro}

Over the past 15 years, it has been discovered that there is a tight correlation between the star formation rate (SFR) of a star-forming galaxy and its existing stellar mass ($M_*$) at any fixed redshift over a wide range of redshifts and environments.  This relationship has been termed the star-forming ``main sequence'' \citep{Noeske2007,Peng2010,Speagle2014,Steinhardt2014a,Schreiber2018}, and is a key constraint on models for galaxy evolution.  In particular, the main sequence suggests that even though an ensemble of star-forming galaxies might live in a range of different environments, with different star formation histories, supernova rates, merger rates, morphology, metallicity, and AGN activity, those differences are observed to have negligible effect on their star formation rates.  Either their impact is minimal or star-forming galaxies are more similar to each other than previously believed. 

Nearly every measurement of the star-forming main sequence has come from a photometric survey.  Templates are then fit to determine properties including stellar masses and star formation rates.  This general technique has been applied in a variety of ways.  Studies of the main sequence have used data from different parts of the electromagnetic spectrum, different star formation rate (SFR) indicators and selection techniques to choose star-forming galaxies, and even different template-fitting techniques to find the best-fit parameters (cf. \citealt{Speagle2014}).  All of these techniques agree on not just the existence of a main sequence, but quantitatively on its location at all $z \lesssim 6$.

However, the main sequences as reported by individual studies initially appeared to disagree.  For example, individual measurements of the $z \sim 2$ main sequence slope $\frac{d\log\textrm{SFR}}{d\log M_*}$ ranged from 0.4 to 0.9.  These differences were shown to be caused by a difference in assumptions about the stellar initial mass function (IMF), dust, and cosmological parameters \citep{Speagle2014}. 

Both components of the main sequence, SFR and $M_*$, depend strongly on the shape of the IMF.  For (stellar) main sequence stars, $L \propto M^{3.5}$, and typical IMFs have $n(M) \propto M^{-2.35}\textrm{ to } M^{-1.3}$\citep{Salpeter1955,Kroupa2001,Chabrier2003}.  Thus, the light is dominated by high-mass stars, but the mass is dominated by low-mass stars.  For star-forming galaxies, the light primarily comes from short-lived O and B stars, whereas for older stellar populations it comes from the most massive stars still on the main sequence.  In both cases, the IMF is used to transform measurements dominated by these relatively rare stars into inferred masses of the very young (used to determine the SFR) and full ($M_*$) stellar population.

In previous studies of the star-forming main sequence, Salpeter \citep{Salpeter1955}, Kroupa \citep{Kroupa2001}, and Chabrier \citep{Chabrier2003} IMFs have been used.  All of these are attempts to describe star formation in the Milky Way.  However, the IMF should depend upon the gas temperature in star-forming regions \citep{LyndenBell1976,Jermyn2018}.  Although Galactic background temperatures are $\sim 20$K \citep{Schnee2008}, dust temperatures in star-forming galaxies are typically greater than 20K \citep{Magnelli2014,Magdis2017,Casey2012}.  Further, at $z > 6.3$, the cosmic microwave background temperature (CMB) also exceeds 20K.  Therefore, it might be expected that the IMF in most star-forming galaxies is different than those derived from Galactic observations.  Such a temperature dependence might even support a feedback mechanism with a main sequence-like attractor solution \citep{Steinhardt2020a}.  

Given the strong dependence of inferred SFR and $M_*$ on the IMF, a different IMF in typical star-forming galaxies can significantly alter the observed main sequence.  Here, these effects are investigated with the use of a new set of photometric templates, described in \citet{Sneppen2022}.  These templates add one new parameter, the IMF `temperature' $T_{IMF}$, expanding the parameter space used in previous studies.  $T_{IMF}$ is translated into an IMF using the prescription in \citet{Jermyn2018}.

In \S~\ref{sec:methods}, the COSMOS2015 catalog, template fitting procedure, and selection of star-forming galaxies are described.  As in \citet{Sneppen2022}, most galaxies are best fit with one of two specific $T_{IMF}$.  This produces a modified main sequence as detailed in  \S~\ref{sec:mainseq}.  The possibility of a connection between the $T_{IMF}$ of massive star-forming galaxies and quiescent galaxies is explored in \S~\ref{sec:quiescent}.  Finally, the implications of these results for proposed evolutionary models are discussed in \S~\ref{sec:discussion}.

This work is Paper II in a series of three related papers.  Paper I discusses the methodology used to measure $T_{IMF}$, along with uncertainty estimates and covariances, and Paper III focuses on quiescent galaxies and quenching.

Analysis presented here uses a flat $\Lambda$CDM cosmology with $(h, \Omega_m, \Omega_\Lambda) = (0.674, 0.315, 0.685)$ \citep{Planck2018} throughout.

\section{Methodology and Temperature-Dependence of the IMF}
\label{sec:methods}

\subsection{Overview}
The methodology used in this work follows the techniques described in Paper I \citep{Sneppen2022}.  These are applied to the COSMOS2015 catalog, which includes broad-band photometry in NUV, u, B, V, r, i, z, Y, J, H, and IRAC channels 1 and 2, two narrow band filters (NB711 and NB816), and 12 intermediate bands, as detailed in \citet{Laigle2016}.  Some objects are not covered by every filter.  In addition, a far more stringent signal-to-noise cut is required in order to constrain the IMF of mock spectra in addition to all other standard parameters.  Therefore, a V-band SNR-cut of 10 is used, with 139,535 galaxies passing this cut \citep{Sneppen2022}. Ultimately, the choice of V-band is arbitrary with high-quality cuts in other bands yielding similar tight-constraints on mock-spectra.

This catalog is then fit with a procedure as similar as possible to established photometric template fitting techniques, with the sole exception that an additional parameter is added to allow variability in the IMF, as described in Paper I.  Using the best-fit redshift, galaxies are separated into star-forming and quiescent populations using rest-frame colors (\S~\ref{uvj-desc}).  Galaxies selected as star-forming are then used to construct the star-forming main sequence.

\subsection{Computing Galactic Parameters}
\label{gal-params}

The photometric template fitting approach used here is based on the Easy and Accurate \zp\ from Yale (EAZY; \citealt{Brammer2008}) software, which has been shown to be successful in photometric redshift determination when compared against other software instruments \citep{Hildebrandt2010}. The analysis yields consistent results when using either the original EAZY implementation in C or the corresponding python wrapper EAZY-py.  In this work, all figures are constructed using the original EAZY source code.  EAZY fits the observed photometric SEDs with a linear combination of 12 basis templates.  These basis templates are themselves representative linear combinations of 560 individual synthetic templates derived using the Flexible Stellar Population Synthesis (FSPS; \citealt{fsps1}, \citealt{fsps2}) instrument, selected so that the 12 basis templates span the full parameter space of observed galaxies.  Each basis template therefore corresponds to a set of physical parameters.  The inferred physical parameters of a galaxy can therefore be constructed from a weighted combination of the parameters corresponding to the 12 basis vectors.

In practice, 12 basis templates can only span a limited portion of the full parameter space provided by the 560 FSPS templates.  The standard library for EAZY chooses this basis through a slightly modified version of non-negative matrix factorization \citep{Brammer2008}, attempting to find basis templates which both span as much of the full space as possible and also correspond to observed galaxy spectra.  The same approach is adopted here for each choice of $T_{IMF}$.  One consequence is that the modified version of EAZY will produce better constraints on $T_{IMF}$ for galaxies which are have many observed analogues (and thus lie close to the space spanned by this basis) than for extreme outliers.

As a simplifying assumption to reduce the size of the parameter space and avoid degeneracies, galaxies are approximated as being dominated by a single IMF, and therefore a single $T_{IMF}$.  Although different stellar sub-populations within a galaxy may well develop with different IMFs, the hope is that these differences average out, so that the population is well-described by a best-fit luminosity-averaged IMF, much in the same way that stellar populations are often described as having a single age.

Fits are performed over a grid of IMFs, spaced every 1K for $8\textrm{K} \leq T_{IMF} \leq 60\textrm{K}$.  In practice, very few objects are fit either below 20K or near the upper bound \citep{Sneppen2022}, so the results here are not sensitive to the choice of cutoff.  A very small population appears to be best fit with temperatures significantly greater than 60K, and is discussed in detail in Paper III.  For each IMF, a set of 560 FSPS templates is constructed corresponding to the same combinations of age, star formation history, extinction, and metallicity as in the standard EAZY library.  Those are then reduced to 12 basis templates, again using the same procedure as for the standard EAZY library.

At each temperature, EAZY is run to compute a best-fit linear combination of the templates to the observed photometry, a goodness of fit expressed as a reduced ${\chi}^2$, and a best fit photometric redshift $z_{phot}$.  The best-fit temperature is the location of the minimum ${\chi}^2$.  For all but 18\% of objects with sufficient signal-to-noise, there is only a single local minimum within the $T_{IMF}$ grid, which is inferred to be the global minimum. The few remaining objects either have multiple local minima or only have a local minimum at 8K or 60K.  A closer examination of these cases revealed that they typically have poorly-constrained photometry, and that additional bands would result in a best-fit temperature between 8K and 60K \citep{Sneppen2022}.  These objects are therefore discarded from the final catalog. 

This minimum corresponds to the best-fit $T_{IMF}$.  Inferred parameters such as $M_*$ and SFR are computed as a luminosity-weighted sum of the basis vectors at that $T_{IMF}$, along the lines of the standard EAZY parameter reconstruction.

\subsection{Filtering Star-Forming and Quiescent Galaxies}
\label{uvj-desc}

\begin{figure}[htbp]
\begin{center}
\includegraphics[width=.95\linewidth]{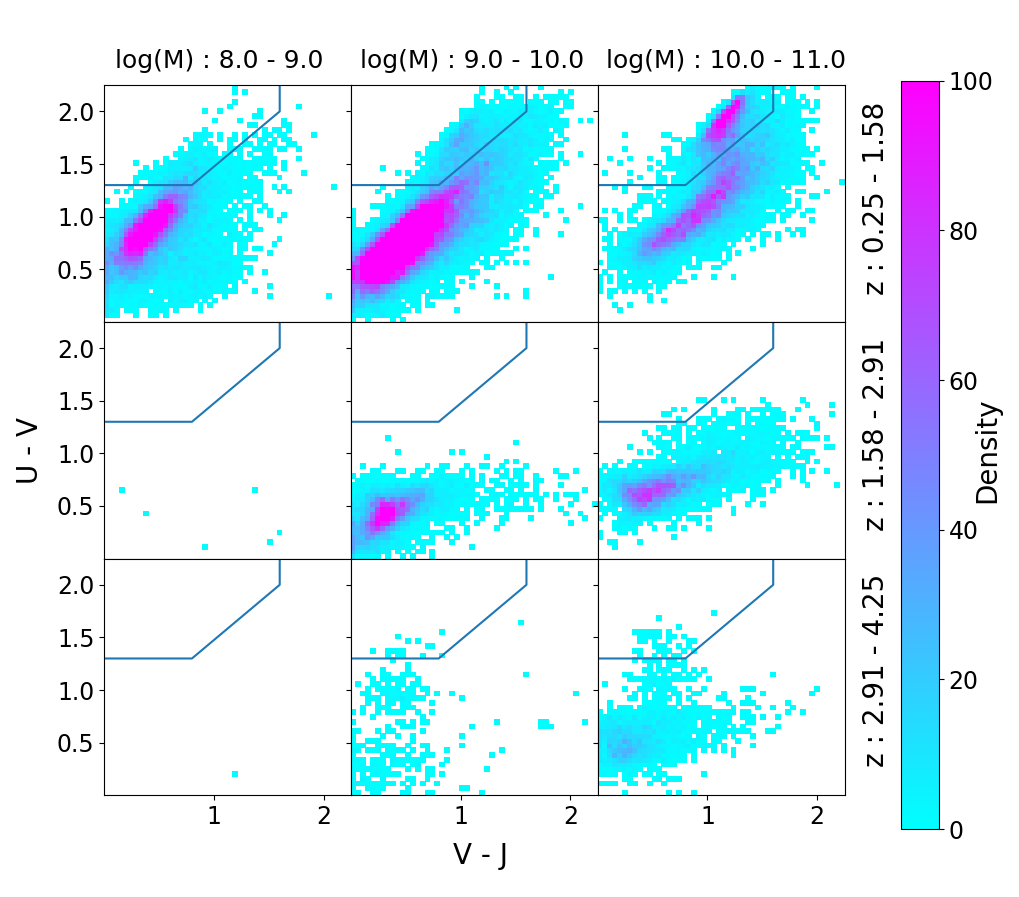}
\caption{UVJ diagram binned by mass and redshift, colored by density.  Quiescent galaxies should lie above and to the left of the indicated boundary, and star-forming galaxies comprise the remainder of the sample.  As reported in numerous previous studies with the COSMOS2015 catalog, galaxies towards higher mass at fixed redshift and at fixed mass towards lower redshift are more likely to be quiescent.}
\label{fig:uvj}
\end{center}
\end{figure}

For the comparisons shown in this work, star-forming and quiescent galaxies are separated primarily using a UVJ diagram \citep{Labbe2005,Brammer2011}. Rest-frame U, V, and J fluxes are calculated from the best-fit reconstructed spectrum, shifted to the rest-frame using the calculated $z_{phot}$ and integrated over the relevant filters. This produces a bimodal distribution (Fig. \ref{fig:uvj}) at low redshifts, similar to previous results. At higher redshifts and lower masses where quiescent galaxies become exceedingly rare as a reflection of downsizing, nearly every galaxy is selected as star-forming.

It is worth noting that the UVJ diagram begins to lose its conventional shape in the lowest mass bin. Since galaxies in the range of $8 < log \frac{M_*}{M_\odot} < 9$ are rare in COSMOS2015 \citep{Furtak2021}, this is likely a reflection of a large population of poorly constrained galaxies being fit with these low masses, perhaps combined with selection effects.

It is known that UVJ selection is incomplete, and will misclassify a few percent of the sample, often due to classifying dusty star-forming galaxies as quiescent. Several approaches have been proposed to solve this problem, including different color selection \citep{Arnouts2013}, template-determined specific star formation rate (sSFR = SFR/$M_*$; \citealt{Brammer2008,Laigle2016}), and even machine learning \citep{Steinhardt2020b}.  As described in \S~\ref{sec:quiescent}, the work presented here offers an additional possible classification.  

For ease of comparison with previous COSMOS2015 main sequence measurements, here a standard UVJ selection is initially used.  In order to produce higher purity samples of star-forming and quiescent galaxies, sSFR is used as an additional screen.  The star-forming sample consists of the top 95\% of galaxies in sSFR from the sample identified by standard UVJ selection \citep{Williams2009}.  Similarly, the quiescent sample consists of the lower 95\% of galaxies in sSFR from the sample identified by standard UVJ selection.  This corresponds to limiting star-forming galaxies to those with $\log \textrm{sSFR} > -9.71$, and limiting quiescent galaxies to those with $\log \textrm{sSFR} < -8.82$, in addition to requiring that they occupy the proper coordinates in UVJ color-space.
The remaining objects are not included in either sample.

\subsection{The Meaning of IMF ``Temperature''}
\label{sec:tempdef}

The family of IMFs used approximates the IMF arising from stars forming out of an isothermal gas at some ambient temperature $T_{IMF}$.  Two temperature-dependent mass scales are relevant: the Jeans mass \citep{Jeans1902} and an adiabatic fragmentation mass \citep{LyndenBell1976}.  \citet{Jermyn2018} associate these with a Kroupa IMF \citep{Kroupa2001} to derive the resulting temperature-dependent IMF as
\begin{equation} \label{eqn:imf-temp}
    \xi(m) \propto 
    \begin{cases} 
      m^{-0.3} & m < 0.08M_\odot \cdot (\frac{T_{IMF}}{T_0})^2 \\
      m^{-1.3} & 0.08M_\odot \cdot (\frac{T_{IMF}}{T_0})^2 < m < 0.5M_\odot \cdot (\frac{T_{IMF}}{T_0})^2 \\
      m^{-2.3} & m > 0.5M_\odot \cdot (\frac{T_{IMF}}{T_0})^2,
    \end{cases}
\end{equation}

where the typical Milky Way gas temperature $T_0$ is assumed to be 20K \citep{Schnee2008,Papadopoulos2010,Steinhardt2020a}.  At $T_{IMF} = 20$K, this reproduces the standard Kroupa IMF.  

Different theoretical and numerical studies have argued for a range of temperature dependences with the mass scaling as $T$ \citep{imf-t}, $T^{3/2}$ \citep{Jeans}, $ T^{2}$ \citep{Steinhardt2020a}, or $T^{5/2}$ \citep{imf-t52}.  Without a second, non-Galactic measurement to calibrate against, the correct temperature dependence cannot be determined merely from template fitting.  The same IMF would be produced by 34 K with a Jeans temperature dependence and 30 K with a Jermyn-Steinhardt temperature dependence. In this work, all temperatures are given in terms of a Jermyn-Steinhardt IMF (Eq. \ref{eqn:imf-temp}). 

It should additionally be noted that because the IMF is derived from fitting the existing stellar population, $T_{IMF}$ does not indicate the gas temperature in star-forming clouds at the time the observed light was emitted.  Rather, $T_{IMF}$ describes gas temperature at the (luminosity-weighted) time the existing stellar population was formed.  For a star-forming galaxy, this might be only 100 Myr before emission, but for a quiescent galaxy it could be several Gyr or more.  Thus, $T_{IMF}$ cannot be compared directly with more instantaneous measurements such as dust temperatures.

\section{The Modified Main Sequence}
\label{sec:mainseq}

\begin{figure}[htbp]
\begin{center}
\includegraphics[width=.95\linewidth]{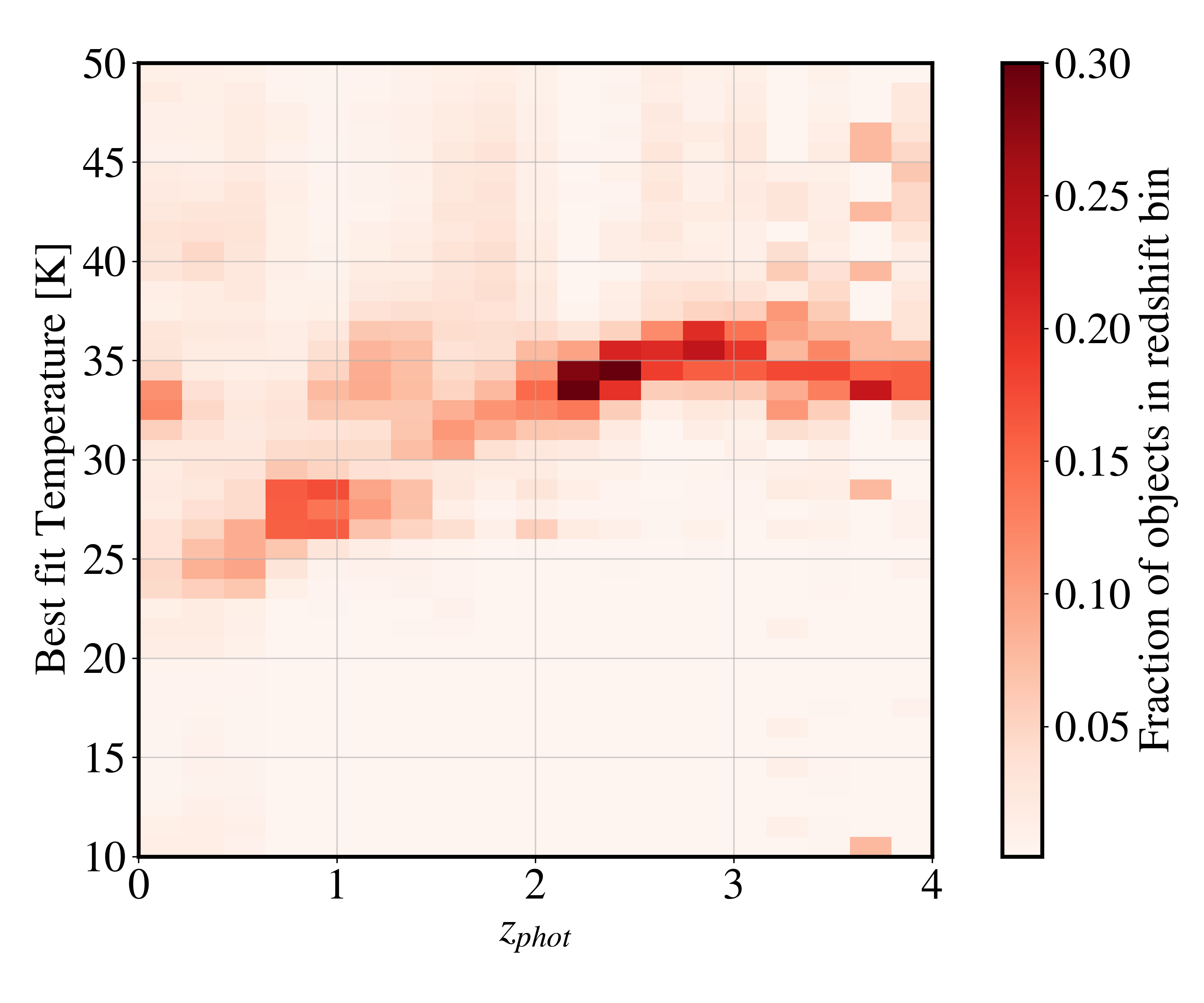}
\caption{Distribution of best-fit $T_{IMF}$ as a function of photometric redshift for star-forming galaxies.  At each redshift, the distribution is individually normalized, in order to emphasize the temperature distribution at each redshift.  The COSMOS2015 catalog contains far fewer galaxies at high redshift than at low redshift, as summarized in \citet{Laigle2016}.  Rather than populating a continuous range of $T_{IMF}$, star-forming galaxies appear to cluster around a specific value at fixed redshift.  Towards higher redshift, $T_{IMF} \sim 35$K, for a bottom-lighter (equivalently, top-heavier) IMF than the Milky Way.  Towards lower redshifts, this characteristic $T_{IMF}$ decreases and the resulting IMF is more similar to a Galactic one, which would correspond to $T_{IMF} = 20$K.} 
\label{fig:sftemps}
\end{center}
\end{figure}

The resulting distribution of best-fit $T_{IMF}$ as a function of redshift is shown in Fig. \ref{fig:sftemps}.  
Most star-forming galaxies at any fixed redshift are best fit with $T_{IMF}$ lying at a characteristic value.  Towards $z = 4$, this approaches 35K, corresponding to a significantly top-heavier IMF than the Galactic one which has previously been assumed.  At lower redshifts, this characteristic temperature decreases, with the best-fit IMF possibly approaching a standard Kroupa IMF ($T_{IMF}= 20$K) towards $z = 0$.

Both stellar mass and SFR are sensitive to the IMF, so with top-heavier IMFs, both quantities differ from the COSMOS2015 catalog.  If all other parameters were fixed, a top-heavier IMF results in both lower $M_*$ and SFR.  However, because extinction, age, and metallicity can also change, and are partially degenerate with $M_*$ and SFR, other behavior is possible.  A summary of these degeneracies and the resulting uncertainties in stellar mass and SFR is given in Paper I \citep{Sneppen2022}, concluding that uncertainties in both quantities are comparable to those in the \citet{Laigle2016} catalog.  The resulting effects on mass functions and the implications for high-redshift mass growth and quiescence are explored in more detail in Paper III \citep{Steinhardt2022b}.  The discussion here focuses on the consequences of these top-heavier IMFs and corresponding IMF temperatures on the star-forming main sequence.

\subsection{Best-fit Evolution}

\begin{figure*}
  \includegraphics[width=.98\linewidth]{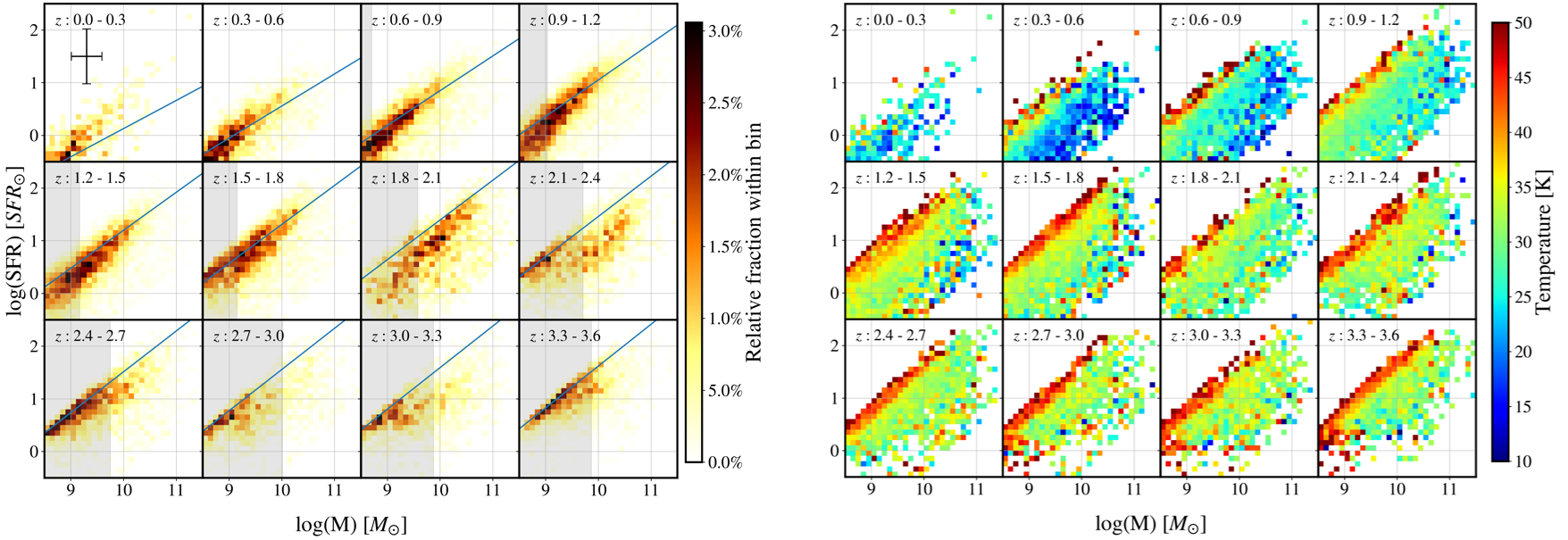}
\caption{Left: Hess diagram of the distribution of stellar mass and star formation rates in bins of $z_{phot}$, with the number density scale normalized to the total number of galaxies in each redshift bin. As in previous studies assuming a Milky Way-like IMF, there continues to be a strong correlation between $M_*$ and SFR characteristic of the star-forming main sequence.  For comparison, the \citet{Speagle2014} main sequence fits are shown in blue.  The grey shading indicates masses below the mass completeness limit with the errorbar in the topleft subplot illustrative of typical uncertainty of stellar mass and star formation rate. Right: $z_{phot}$-binned star-forming main sequence colored by temperature. Star-forming galaxies at a given redshift tend to become cooler at higher masses at fixed redshift, and exhibit a temperature gradient across the star-forming main sequence, with hotter galaxies lying at higher SFRs for a given mass.} 
\label{fig:temp-z-all}
\end{figure*}

The new stellar masses and SFRs computed following the methodology in \S~\ref{gal-params} continue to exhibit a tight correlation at fixed redshift (Fig. \ref{fig:temp-z-all}), a behavior which has been termed the star-forming main sequence.

The location and slope of the main sequence has remained largely unchanged, with a slightly higher slope at low redshift than previous measurements (blue lines).  To first approximation, the best-fit bottom-lighter IMF for star-forming galaxies reduces both stellar mass and SFR, but produces a similar relationship.

Although a qualitative comparison is possible, the best-fit linear SFR($M_*$) depends upon the uncertainties in both SFR and $M_*$ as well as mass completeness.  Because the IMF is fit using a grid, whereas other parameters are then fit in continuous manner at each fixed IMF, the fit uncertainties produced by EAZY do not represent the true uncertainties in SFR and $M_*$.  It would be possible to estimate those uncertainties using a grid of parameters around the best-fit solution, but doing so will require using grid-based template fitting rather than EAZY.  As a result, a quantitative best-fit main sequence is beyond the scope of this work.

\subsection{Temperature Evolution Along the Main Sequence}

It is also natural to consider the evolution of the IMF, and of $T_{IMF}$, both along the main sequence and with redshift.  In general, star-forming galaxies have higher $T_{IMF}$ towards higher redshifts (Fig. \ref{fig:temp-z-all}).  This might be expected for any of several reasons.  Star formation rates are higher, and a young stellar population heats gas both due directly to radiated light while on the main sequence and less directly due to cosmic rays produced in supernovae \citep{Papadopoulos2010,Steinhardt2020a}.  Higher gas densities, both due to more gas \citep{Santini2014} and smaller radii \citep{Mowla2019}, allow more efficient cooling, and thus can sustain higher temperatures in equilibrium.  Finally, at very high redshift, even the CMB contribution becomes relevant \citep{Jermyn2018}.

Perhaps more surprising is the complex and heterogeneous temperature distribution of the main sequence at fixed redshift.  At fixed stellar mass, a higher SFR is associated with higher $T_{IMF}$, an effect which was also previously reported for dust temperatures \citep{Magnelli2014}.  There is also a narrow region at very high redshift with $T_{IMF}$ well above the 35-40K which is typical of hot star-forming galaxies.  There is likely a strong overlap between this population and ultra-luminous infrared galaxies (ULIRGs; \citealt{Lonsdale2006}), which will be explored in future work. 

For the bulk of the main sequence, the average $T_{IMF}$ decreases slightly with increasing mass.  In particular, for the most massive star-forming galaxies, which should be closest to turning off, the star-forming main sequence also has the lowest $T_{IMF}$.  Perhaps this is an indication that an early stage in quenching involves lowering gas temperatures in star-forming regions, or in some other way altering the IMF of newly-formed stars.  A connection between this idea and the $T_{IMF}$ measured for quiescent galaxies is discussed in \S~\ref{sec:quiescent}.  

Another possibility is that $T_{IMF}$, which should approximate the gas temperature in star-forming regions, is a measure of the cool gas density.
Since denser gas can radiate more efficiently, it can sustain higher equilibrium temperatures.  Thus, a galaxy with a greater fraction of molecular hydrogen, and thus presumably denser molecular clouds, might have both higher sSFR and $T_{IMF}$.  Since the star-forming main sequence has a slope less than 1, the same low-mass galaxies which have higher $T_{IMF}$ will also have higher sSFR.  Similarly, sSFR increases at fixed mass towards higher redshift, as does $T_{IMF}$.

\subsection{Comparing Gas and Dust Temperatures}

It it natural to compare the gas temperatures described here with other indicators.  The best established is dust temperature, which can be measured in several different ways.  Dust can have multiple components, and it is likely that the gas in the coolest, star-forming molecular clouds should be most closely in equilibrium with the coolest dust component.  

Indeed, measurements of dust temperatures in star-forming galaxies also find typical temperatures in the 25-40K range.  \citet{Magnelli2014} use a stacked analysis to compute the average $T_{dust}$ along the star-forming main sequence, and their Fig. 6 can be directly compared with Fig. \ref{fig:temp-z-all} in this work.  

Dust temperatures display two of the three behaviors seen in $T_{IMF}$, but not the third.  There is an increase in both the average $T_{dust}$ and $T_{IMF}$ towards high redshift.  Similarly, at fixed stellar mass and redshift, both temperatures increase with increasing SFR.  However, at fixed redshift, the center of the star-forming main sequence is approximately isothermal in $T_{dust}$, whereas $T_{IMF}$ decreases towards high mass.  If $T_{IMF}$ and $T_{dust}$ truly measure gas and dust temperatures, this difference might allow a useful diagnostic of the gas-dust relationship in star-forming galaxies.  Possible explanations for this are considered in \S~\ref{sec:discussion}.

Still, the similarities between $T_{dust}$ and $T_{IMF}$ suggest that they may indeed be in near-equilibrium.  Indeed, one might wonder whether it is possible to use $T_{dust}$ measurements to calibrate $T_{IMF}$.  Different theoretical and numerical studies have all associated higher temperatures with top-heavier IMFs, but with different temperature dependence (\S~\ref{sec:tempdef}).  Perhaps dust temperatures could be used to select the proper scaling.

Unfortunately, due to the low resolution of far-infrared data, it is often the case that dust temperatures of high-redshift galaxies are poorly constrained. This is due to degeneracy between temperature and other parameters, especially the emissivity spectral index $\beta$ (as in Eqs \ref{eq:mbb}-\ref{eq:mbbpl}). If $\beta$ is fixed at an assumed value, then temperature appears to be better constrained ($1\sigma \sim 1K$), but if $\beta$ is not assumed (as there is little physical motivation to do so), both temperature and $\beta$ are poorly constrained ($1\sigma \sim 10K$) in a predictable manner. Furthermore, there exist several dust temperature models, none of which reliably outperform than the others. The results of these models often differ systematically and enormously. For these reasons, with current data and models one cannot determine with sufficient certainty the dust temperatures of high-redshift galaxies.

As an example, here seven models chosen for their use in recent studies are fit to objects in the COSMOS catalog: A modified black-body (MBB; \citealt{Casey2012}), 
\begin{equation}
\label{eq:mbb}
B_\nu (\nu, T) = N_{bb} \frac{\nu^3 (1-e^{-\left(\frac{\lambda_0}{\lambda}\right)^\beta})}{e^{h\nu /(k_B T)} - 1};
\end{equation}
a modified black-body approximated as optically thin (MBB + OT), 
\begin{equation}
\label{eq:mbbot}
B_\nu (\nu, T) = N_{bb} \frac{\nu^{3+\beta}} {e^{h\nu /(k_B T)} - 1};
\end{equation}
a modified black-body curve with an additional power law (MBB + PL) to approximate contributions from higher-temperature bodies, 
\begin{equation}
\label{eq:mbbpl}
B_\nu (\nu, T) = N_{bb} \nu^3 \frac{1-e^{-\left(\frac{\lambda_0}{\lambda}\right)^\beta}}{e^{h\nu /(k_B T)} - 1} + N_{pl} \lambda^\alpha e^{-(\lambda/\lambda_c)^2},
\end{equation}
with three choices of power-law index $\alpha \in \{1,2,3\}$; an optically-thin modified black-body with a power-law contribution (MBB + PL + OT), and the \citet{Draine2007} model with $\gamma = 0.02$ (DL07). These models were fit using least-squares regression on photometric data in the $\sim$ .1 to 1 mm bands.

A comparison of these models on the 194 COSMOS objects for which all are well-fit (Table \ref{tab:dust}) finds that dust temperature estimates vary substantially between different models, with systematic differences that far exceed their statistical uncertainty. The statistical uncertainty of these fits was typically $\lesssim 1K$, and thus the differences in these models can be attributed nearly entirely to systematic uncertainty both for individual objects and for the ensemble average.  As a result, it is difficult to compare the gas temperatures estimated in this work with dust temperatures, except to conclude that they lie in a similar range.

\begin{table}
\caption{Average temperatures resulting from} fitting seven different dust models to the COSMOS catalog. This table displays data about fits on only a set of galaxies which all the models were able to fit with a $\chi^2_r <10$ ($n=194$).  Dust temperature estimates vary substantially between different models.  As a result, it is difficult to compare gas temperatures estimated in this work with dust temperatures.\label{tab:dust}
\begin{tabular}{lllll}
    \cline{2-5}
                    & $\beta$              & 1.5   & 1.8   & 2.0   \\\cline{2-5}
    MBB           & \mean{T}             & 35.42 & 35.78 & 35.95 \\
                      & \mean{\chi^2_r}      & 1.255 & 0.963 & 0.855 \\  \cline{2-5} 
    MBB + OT      & \mean{T}             & 26.64 & 24.80 & 23.71 \\
                      & \mean{\chi^2_r}      & 1.434 & 1.482 & 1.568 \\  \cline{2-5}
        
    MBB + PL      & \mean{T}             & 35.40 & 35.44 & 35.68 \\
        $\alpha = 1$  & \mean{\chi^2_r}      & 1.239 & 0.953 & 0.888 \\ \cline{2-5}
        
    MBB + PL      & \mean{T}             & 30.88 & 31.24 & 30.34 \\
        $\alpha = 2$  & \mean{\chi^2_r}      & 1.523 & 1.183 & 1.142 \\ \cline{2-5}
        
    MBB + PL      & \mean{T}             & 33.37 & 33.05 & 33.05 \\
        $\alpha = 3$  & \mean{\chi^2_r}      & 2.262 & 1.949 & 1.810 \\ \cline{2-5}
        
    MBB + PL + OT & \mean{T}             & 23.34 & 20.47 & 20.44 \\
        $\alpha = 2$  & \mean{\chi^2_r}      & 1.769 & 1.820 & 1.665 \\ \cline{2-5}

                      & $q_{PAH}$            & 0     & 1     & 2     \\ \cline{2-5}
    DL07          & \mean{U_{min}}       & 19.30 & 14.08 & 10.47 \\ 
        $\gamma = .02$& \mean{\chi^2}      & 8.453 & 9.456 & 10.61 \\  \cline{2-5} 
        \end{tabular}
\end{table}

\subsection{Cosmic Microwave Background Contribution}

The CMB temperature increases as $(1+z)$, reaching the Milky Way $T_{IMF}$ of 20K at $z = 6.3$.  CMB photons should be a source of heating for the entire galaxy, as long as there are minimal amounts of dust to maintain equilbrium \citep{Jermyn2018}.  Thus, it should be expected that for galaxies at $z \gtrsim 6$, star-forming clouds must be warmer than in the Milky Way, resulting in a top-heavier IMF.  Indeed, a search for this effect at high redshift was one of the rationales for developing the templates used in this work in preparation for upcoming high-redshift observations with the James Webb Space Telescope (JWST).  

However, the best-fit $T_{IMF}$ for high-redshift galaxies is not the 20K of the Milky Way, but rather closer to 35-40K.  Although at low redshift there is a population of cooler quiescent galaxies, at very high redshift it is expected that essentially every bright galaxy must still be actively star-forming.  Thus, the redshift at which the CMB becomes relevant should be closer to $z \gtrsim 12-14$, and possibly higher if a denser galaxy with higher sSFR results in increased $T_{IMF}$ towards high redshift.  This is a higher redshift than any galaxy currently known, and may even be higher than any galaxy found in planned JWST surveys.  

If at some redshift the CMB provides a meaningful lower bound on $T_{IMF}$, it would be detected as a linear increase in $T_{IMF,min}(1+z)$ above some minimum $(1+z)$.  This would also potentially be a way of calibrating the relationship between $T_{IMF}$ and gas temperature (cf. \S~\ref{sec:tempdef}).  Because $T_{IMF}$ describes the gas at the time of star formation rather than observed light emission, perhaps a $z \sim 10$ galaxy with an older stellar population might still allow a detection of this effect.  However, for most galaxies even at very high redshift, the effects of the CMB on the IMF are unlikely to be detectable by current and upcoming facilities.

\section{Quiescent Galaxies}
\label{sec:quiescent}

It is well known that galaxies can be divided into two broad classes, star-forming galaxies and quiescent galaxies.  These distributions are clearly distinct in color-color diagrams such as UVJ \citep{Williams2009} and NUVrJ \citep{Ilbert2013}, and as the names imply, are comprised of galaxies with very different SFR, sSFR, and physical properties. 

\begin{figure}[htbp]
\begin{center}
\includegraphics[width=.95\linewidth]{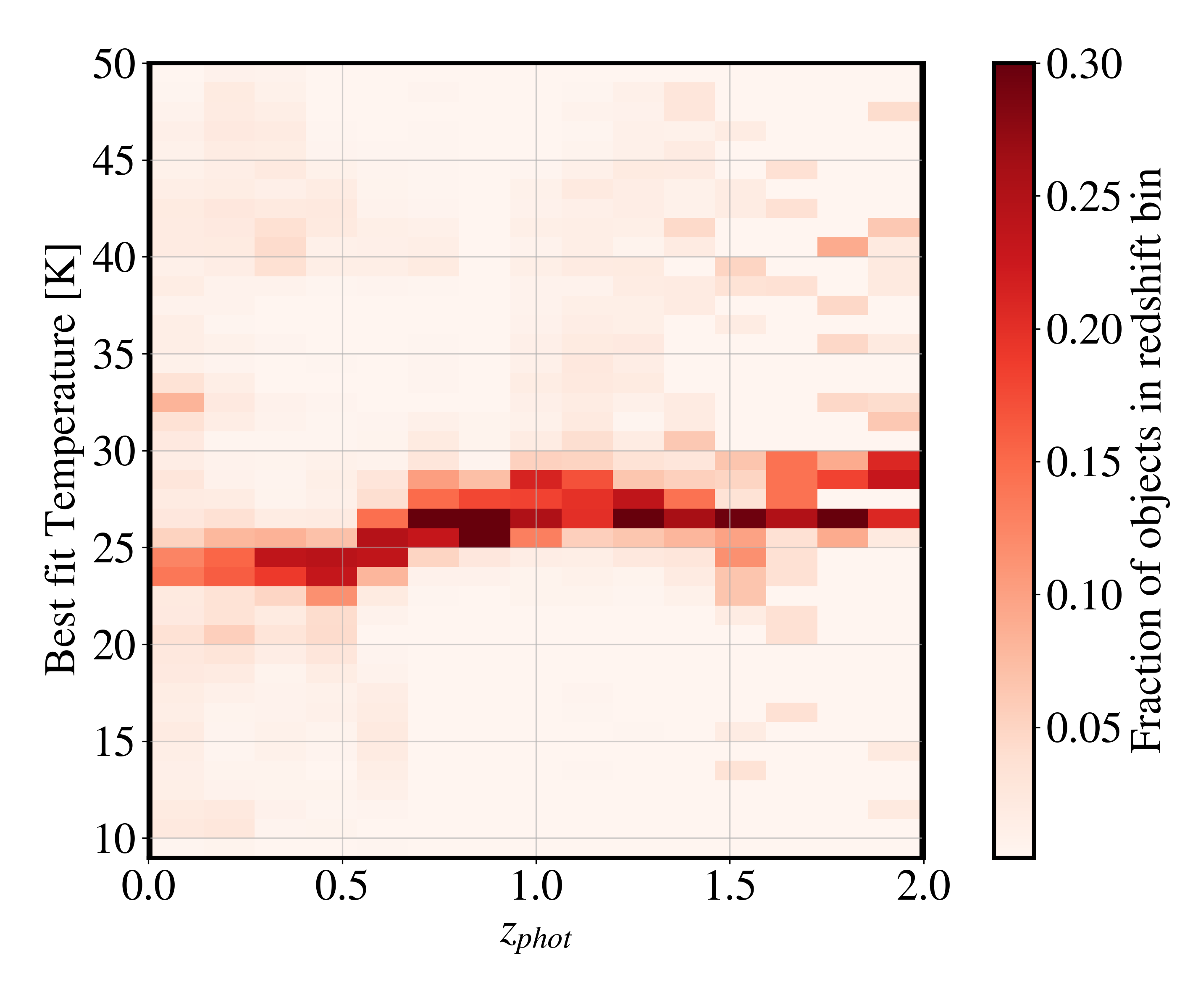}
\caption{Distribution of best-fit $T_{IMF}$ as a function of photometric redshift for quiescent galaxies. Quiescent galaxies cluster around a typical $T_{IMF}$ at each redshift, with IMF temperatures ranging from 25-30K, perhaps slightly increasing with redshift up to $z \sim 2$. Very few quiescent galaxies are found at higher redshifts.  These $T_{IMF}$ are similar to those of star-forming galaxies at $z \sim 0$, but lower at higher redshifts.}
\label{fig:quiescenttemps}
\end{center}
\end{figure}

The distribution of best-fit $T_{IMF}$ also appears to differ between the two classes.  For both populations, galaxies appear to exhibit a characteristic $T_{IMF}$ at every fixed redshift.  However, whereas the star-forming $T_{IMF}$ increases towards higher redshift (Fig. \ref{fig:sftemps}), the quiescent $T_{IMF}$ is nearly constant, perhaps exhibiting a slight increase from $z = 0$ to $z \sim 2$ (Fig. \ref{fig:quiescenttemps}).

Towards $z = 0$, the best-fit IMFs for star-forming and quiescent galaxies are similar.  However, at most redshifts, star-forming galaxies exhibit higher $T_{IMF}$.  A possible hint at an explanation lies in the observation that the most massive galaxies on the star-forming main sequence are typically also coolest (Fig. \ref{fig:temp-z-all}), having $T_{IMF}$ most similar to quiescent galaxies.  It has been known for a few decades that the most massive galaxies become quiescent earlier than less massive galaxies, one of several processes that have been labeled `downsizing' \citep{Cowie1996,Fontana2006,Stringer2009,Fontanot2009}.  If the most massive star-forming galaxies at each redshift are in the early stages of quenching, then perhaps the shift to a lower $T_{IMF}$ more like that of quiescent galaxies provides an indicator.

This would be particularly useful because the selection of quiescent galaxies has a significant time delay.  A color-color selection, such as UVJ or NUVrJ, typically only identifies galaxies which no longer have the massive and luminous blue stars which dominate the light from a young stellar population but have shorter lifetimes.  However, while those stars are still luminous, the colors of a star-forming and recently quenched galaxy will be nearly indistinguishable.  It likely takes $0.5 - 1$ Gyr after star formation stops for a galaxy to be selected as quiescent \citep{Wild2020}.  Star formation rates in template fitting are also dominated by this young population, and provide an average SFR over $\sim 10^8$ years rather than an instantaneous measurement.

The most intriguing possibility would therefore be that a drop in $T_{IMF}$ might provide a significantly earlier indicator of quenching.  $T_{IMF}$ is already backward-looking because it measures the gas conditions around the time the stellar population formed rather than when the observed light was emitted.  If the early stages of quenching are associated with lower gas temperatures, $T_{IMF}$ should fall earlier than other indicators based on galaxy colors.  

The backward-looking nature of $T_{IMF}$ also provides an explanation for why the characteristic $T_{IMF}$ could be redshift-dependent for star-forming galaxies but not for quiescent ones.  Since $T_{IMF}$ measures the conditions under which the observed stellar population formed, and quiescent galaxies are not forming new stars, $T_{IMF}$ will be unchanged even if the gas in aging galaxies continue to cool.  However, star-forming galaxies on the main sequence typically have an stellar population age of $\sim 10^8$ yr, so that lower gas temperatures towards lower redshift would be reflected in not just a lower best-fit SFR, but also a lower $T_{IMF}$.

\begin{figure}[htbp]
\begin{center}
\includegraphics[width=\linewidth]{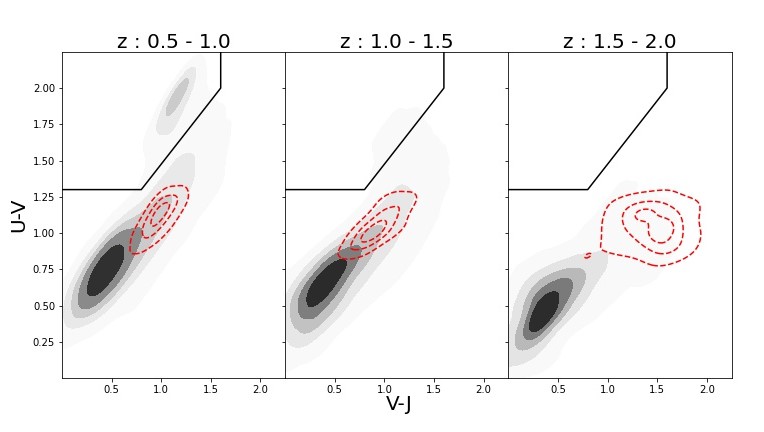}
\caption{UVJ diagram comparing the distribution of all objects in three different redshift ranges (grey contours/shading) with the distribution of the high-mass ($M_* > 10^{10} M_\odot$), lowest-temperature ($T_{IMF} < 27$ K) objects in each redshift panel (red contours).  This sub-population are candidates for star-forming galaxies which are in the process of quenching, but have not yet had enough time to change from blue to red.  There is considerable overlap between this sample and the SF3 galaxies with the oldest blue sequences described in \citet{Wild2014}.}
\label{fig:uvjquench}
\end{center}
\end{figure}

Ideally, the same effect might even provide the ability to select quenching galaxies rather than merely ones which have already quenched.  A primary difficulty in testing quenching mechanisms has been an inability to identify galaxies in these earliest stages.  Perhaps selecting galaxies with star-forming colors but quiescent $T_{IMF}$ might produce a suitable population for investigating the mechanisms by which star formation stops in massive galaxies.  Indeed, such galaxies occupy a distinct region on a UVJ diagram (Fig. \ref{fig:uvjquench}).  This region of the UVJ diagram also has considerable overlap with the subpopulation of SF3 galaxies with the oldest blue sequences described in \citet{Wild2014} as possible precursors of quiescent galaxies.

\section{Discussion}
\label{sec:discussion}
The introduction of an additional parameter into photometric template fitting, $T_{IMF}$, allows galaxies to be fit with any of a family of stellar initial mass functions based on the prescription in \citet{Jermyn2018}.  These IMFs have similar shape to a Kroupa IMF, but with the knees, or breakpoints, at higher masses with increasing $T_{IMF}$.

Although some galaxies at very low redshift are best-fit with a Milky Way-like IMF, most galaxies are instead fit with one of two other IMFs.  The bulk of star-forming galaxies are typically fit with a much bottom-lighter (or top-heavier) IMF than the Milky Way, corresponding to an inferred gas temperature of $\sim 35-40$K.  Quiescent galaxies, along with a smaller fraction of star-forming ones, are instead fit with $T_{IMF} \sim 25$K, producing an IMF only slightly bottom-lighter than the Milky Way.

A particularly intriguing feature of these IMFs is that quiescent galaxies nearly all are fit with the lower $T_{IMF}$, while most star-forming galaxies are fit with the higher value.  The star-forming galaxies which do exhibit lower $T_{IMF}$ are on the very high-mass end of the star-forming main sequence, and likely about to turn off and become quiescent.  That is, these galaxies appear star-forming in color space, but have an IMF characteristic of quiescent galaxies and are likely to also appear quiescent in color space in the near future.  Given that there should be a $\sim 0.5-1$ Gyr delay between quenching and sufficient color changes to select a galaxy as quiescent in color space, the most promising explanation is that $T_{IMF}$ provides an earlier indicator of quenching than color selection.  If so, galaxies with star-forming colors but quiescent $T_{IMF}$ might be in the midst of quenching, and followup studies of that population would then yield useful constraints on quenching mechanisms.

Most of the previously-established features of the star-forming main sequence remain with these new IMFs.  There is a still a tight correlation between SFR and $M_*$ at every redshift where it can be tested.  The relationship is still well fit by a power law with exponent $< 1$, meaning that higher-mass galaxies have lower sSFR at fixed redshift.  The sSFR of star-forming galaxies at any fixed mass still increases towards higher redshifts.  However, surprising features also appear in this analysis which may be useful for developing and testing physical models.  

\subsection{Physical Interpretation of Redshift Evolution in IMF "Temperature"}

Perhaps the most striking new feature of these fits is that fixed redshift, most star-forming galaxies are best-fit with a specific IMF, as are most quiescent galaxies.  However, the two exhibit different redshift evolution, with the characteristic star-forming $T_{IMF}$ higher towards high redshift and the characeristic quiescent $T_{IMF}$ nearly redshift-independent.  At most redshifts, $T_{IMF}$ is higher for star-forming galaxies.  

Thus, at any specific redshift, most galaxies are best-fit with one of two distinct IMFs (Figs. \ref{fig:sftemps}-\ref{fig:quiescenttemps}).  This is not what might have been expected intuitively.  Indeed, much of the rationale for this work was the cosmic ray-driven model developed in \citet{Steinhardt2020a}, which was capable of reproducing, at least qualitatively, the key features of the star-forming main sequence.  $T_{IMF}$ would then be set by an equilibrium between the temperature increase due to additional star formation and cosmic ray generation and a resulting decrease in SFR as temperature rises, allowing fewer clouds to condense into new stars.  

A concern in building such a model was that there were enough free parameters to make it difficult, or even potentially impossible, to test.  Galaxies might follow any of a large family of tracks along which they would drop in $T_{IMF}$ at different rates, moving through a continuous family of IMFs.  Although galaxies of the same baryonic mass might lie on similar tracks, the full set of star-forming galaxies at fixed redshift should encompass a variety of different conditions, and therefore a wide range of $T_{IMF}$.  With the results here, the model not only becomes falsifiable, but can be rejected.  Rather, there must be an additional component of feedback which regulates $T_{IMF}$ even across different environments.

However, it is also surprising that there are predominantly two distinct IMFs at fixed redshift.  For example, at $z = 2$, quiescent and star-forming galaxies have estimated gas temperatures of around $25\,\mathrm{K}$ and $35\,\mathrm{K}$, corresponding to upper knee masses of $0.75 M_\odot$ and $1.5 M_\odot$ respectively for a Kroupa-like IMF.  Few galaxies lie either between these modes or with more extreme IMFs.  Rather, there are two sets of typical conditions, one for quiescent galaxies and one for star-forming ones.  So it is natural to wonder what physics limits galaxies of a wide range of masses and ages to just these two distinct options.

Two mechanisms for producing discrete IMFs seem most plausible.  The first is that the physics of collapse and fragmentation is fundamentally different in galaxies with one IMF than in those with the other, and that each mechanism has a characteristic equilibrium.  For example, it could be that in some galaxies turbulent pressure supports molecular clouds against collapse \citep{Hopkins2012} and in others the thermal pressure dominates \citep{Jeans1902}.  In this scenario it is not surprising that distinct physical processes result in different initial stellar masses, which then explains the observed bimodality.

The second approach is to posit that the same physics is at work in galaxies with both $T_{IMF}$, but that the input physical parameters are different.  Indeed, by parameterizing the IMF with the temperature of molecular clouds, it has been assumed here that one effect, thermal pressure support, halts the fragmentation process in all galaxies.  In this picture, there would need to be some mechanism that can only produce two possible mean molecular cloud temperatures, and that this then acts through pressure support to cause the two observed IMFs.

The main challenges for the first option lie in understanding (i) which physical processes are at work in the two different types of galaxy and (ii) what determines which process dominates in each galaxy.  For instance, if the processes are turbulent and thermal pressure-supported, it would be necessary to understand why some galaxies have significantly more turbulence in their molecular clouds than others, and what sets that scale.  This is not the only possible mechanism, however, and a great diversity of processes have been proposed including radiative feedback \citep{Raskutti2016}, magnetic pressure support \citep{Hennebelle2019}, and cosmic ray pressure support \citep{Papadopoulos2010,Steinhardt2020a}.

The second option poses a similar set of questions.  The most essential are which process regulates collapse during star formation, and why it prefers exactly two distinct distributions of initial stellar mass.  If the relevant process is thermal pressure support, then the question sharpens to asking why molecular clouds are hot in some galaxies and cold in others, and what sets those two scales.  At present, the combination of added, free parameters and limited observational constraints makes it difficult to distinguish between these two scenarios.  

An enticing solution here is that the temperature of molecular clouds is set by the incident stellar and cosmic radiation.  Because the production of both starlight and cosmic rays is sensitive to the IMF, it is conceivable that the temperature of molecular clouds in a galaxy co-evolves with its stellar population, in a way that supports multiple attractor states \citep{Steinhardt2020a}.  Then, each galaxy would fall into one of the attractor states and stay there until either kicked out by mergers or some subsequent evolution destabilizes the attractor state.
Unfortunately, as discussed above, toy models with this phenomenology do not produce a bimodal temperature distribution, but rather a bimodal specific star formation rate.

Another possibility comes from different types of dust in the interstellar medium (ISM).  Dust in the ISM falls into multiple discrete types, each produced in different ways \citep{Gall2011,Lesniewska2019}.  For example, iron is primarily synthesized in Type Ia and core collapse supernovae \citep{Matsuura2011}.  Carbon dust can be produced in lower-mass stars, without the need for a supernova \citep{Gail1999,Gomez2012,Sloan2017}.  If $T_{IMF}$ is set by thermal equilibrium involving ISM dust, then an ISM dominated by dust of different composition might yield different equilibria. 

The bimodal distribution of $T_{IMF}$, along with the association of one IMF with quiescent galaxies and possibly the other with star formation, presents a new constraint on feedback models.  It is unclear which of several possible explanations is most promising, but at present, some new mechanism appears to be required to produce an explanation.

The authors would like to thank Vasily Kokorev and Darach Watson for useful discussions.  CLS is supported by ERC grant 648179 "ConTExt".  BM is supported by the Tombrello Fellowship. AL is supported by the Selove Prize.  The Cosmic Dawn Center (DAWN) is funded by the Danish National Research Foundation under grant No. 140. 
The Flatiron Institute is supported by the Simons Foundation.

\bibliographystyle{mnras}
\bibliography{refs.bib} 

\begin{thebibliography}{}
\makeatletter
\relax
\def\mn@urlcharsother{\let\do\@makeother \do\$\do\&\do\#\do\^\do\_\do\%\do\~}
\def\mn@doi{\begingroup\mn@urlcharsother \@ifnextchar [ {\mn@doi@}
  {\mn@doi@[]}}
\def\mn@doi@[#1]#2{\def\@tempa{#1}\ifx\@tempa\@empty \href
  {http://dx.doi.org/#2} {doi:#2}\else \href {http://dx.doi.org/#2} {#1}\fi
  \endgroup}
\def\mn@eprint#1#2{\mn@eprint@#1:#2::\@nil}
\def\mn@eprint@arXiv#1{\href {http://arxiv.org/abs/#1} {{\tt arXiv:#1}}}
\def\mn@eprint@dblp#1{\href {http://dblp.uni-trier.de/rec/bibtex/#1.xml}
  {dblp:#1}}
\def\mn@eprint@#1:#2:#3:#4\@nil{\def\@tempa {#1}\def\@tempb {#2}\def\@tempc
  {#3}\ifx \@tempc \@empty \let \@tempc \@tempb \let \@tempb \@tempa \fi \ifx
  \@tempb \@empty \def\@tempb {arXiv}\fi \@ifundefined
  {mn@eprint@\@tempb}{\@tempb:\@tempc}{\expandafter \expandafter \csname
  mn@eprint@\@tempb\endcsname \expandafter{\@tempc}}}

\bibitem[\protect\citeauthoryear{{Arnouts} et~al.,}{{Arnouts}
  et~al.}{2013}]{Arnouts2013}
{Arnouts} S.,  et~al., 2013, \mn@doi [\aap] {10.1051/0004-6361/201321768},
  \href {https://ui.adsabs.harvard.edu/abs/2013A&A...558A..67A} {558, A67}

\bibitem[\protect\citeauthoryear{{Brammer}, {van Dokkum}  \& {Coppi}}{{Brammer}
  et~al.}{2008}]{Brammer2008}
{Brammer} G.~B.,  {van Dokkum} P.~G.,   {Coppi} P.,  2008, \mn@doi [\apj]
  {10.1086/591786}, \href
  {https://ui.adsabs.harvard.edu/abs/2008ApJ...686.1503B} {686, 1503}

\bibitem[\protect\citeauthoryear{{Brammer} et~al.,}{{Brammer}
  et~al.}{2011}]{Brammer2011}
{Brammer} G.~B.,  et~al., 2011, \mn@doi [\apj] {10.1088/0004-637X/739/1/24},
  \href {https://ui.adsabs.harvard.edu/abs/2011ApJ...739...24B} {739, 24}

\bibitem[\protect\citeauthoryear{{Casey}}{{Casey}}{2012}]{Casey2012}
{Casey} C.~M.,  2012, \mn@doi [\mnras] {10.1111/j.1365-2966.2012.21455.x},
  \href {https://ui.adsabs.harvard.edu/abs/2012MNRAS.425.3094C} {425, 3094}

\bibitem[\protect\citeauthoryear{{Chabrier}}{{Chabrier}}{2003}]{Chabrier2003}
{Chabrier} G.,  2003, \mn@doi [\pasp] {10.1086/376392}, \href
  {https://ui.adsabs.harvard.edu/abs/2003PASP..115..763C} {115, 763}

\bibitem[\protect\citeauthoryear{{Chabrier}, {Hennebelle}  \&
  {Charlot}}{{Chabrier} et~al.}{2014}]{imf-t52}
{Chabrier} G.,  {Hennebelle} P.,   {Charlot} S.,  2014, \mn@doi [\apj]
  {10.1088/0004-637X/796/2/75}, \href
  {https://ui.adsabs.harvard.edu/abs/2014ApJ...796...75C} {796, 75}

\bibitem[\protect\citeauthoryear{{Conroy} \& {Gunn}}{{Conroy} \&
  {Gunn}}{2010}]{fsps2}
{Conroy} C.,  {Gunn} J.~E.,  2010, \mn@doi [\apj]
  {10.1088/0004-637X/712/2/833}, \href
  {https://ui.adsabs.harvard.edu/abs/2010ApJ...712..833C} {712, 833}

\bibitem[\protect\citeauthoryear{{Conroy}, {Gunn}  \& {White}}{{Conroy}
  et~al.}{2009}]{fsps1}
{Conroy} C.,  {Gunn} J.~E.,   {White} M.,  2009, \mn@doi [\apj]
  {10.1088/0004-637X/699/1/486}, \href
  {https://ui.adsabs.harvard.edu/abs/2009ApJ...699..486C} {699, 486}

\bibitem[\protect\citeauthoryear{{Cowie}, {Songaila}, {Hu}  \& {Cohen}}{{Cowie}
  et~al.}{1996}]{Cowie1996}
{Cowie} L.~L.,  {Songaila} A.,  {Hu} E.~M.,   {Cohen} J.~G.,  1996, \mn@doi
  [\aj] {10.1086/118058}, \href
  {https://ui.adsabs.harvard.edu/abs/1996AJ....112..839C} {112, 839}

\bibitem[\protect\citeauthoryear{{Draine} \& {Li}}{{Draine} \&
  {Li}}{2007}]{Draine2007}
{Draine} B.~T.,  {Li} A.,  2007, \mn@doi [\apj] {10.1086/511055}, \href
  {https://ui.adsabs.harvard.edu/abs/2007ApJ...657..810D} {657, 810}

\bibitem[\protect\citeauthoryear{{Fontana} et~al.,}{{Fontana}
  et~al.}{2006}]{Fontana2006}
{Fontana} A.,  et~al., 2006, \mn@doi [\aap] {10.1051/0004-6361:20065475}, \href
  {https://ui.adsabs.harvard.edu/abs/2006A&A...459..745F} {459, 745}

\bibitem[\protect\citeauthoryear{{Fontanot}, {De Lucia}, {Monaco}, {Somerville}
   \& {Santini}}{{Fontanot} et~al.}{2009}]{Fontanot2009}
{Fontanot} F.,  {De Lucia} G.,  {Monaco} P.,  {Somerville} R.~S.,   {Santini}
  P.,  2009, \mn@doi [\mnras] {10.1111/j.1365-2966.2009.15058.x}, \href
  {https://ui.adsabs.harvard.edu/abs/2009MNRAS.397.1776F} {397, 1776}

\bibitem[\protect\citeauthoryear{{Furtak}, {Atek}, {Lehnert}, {Chevallard}  \&
  {Charlot}}{{Furtak} et~al.}{2021}]{Furtak2021}
{Furtak} L.~J.,  {Atek} H.,  {Lehnert} M.~D.,  {Chevallard} J.,   {Charlot} S.,
   2021, \mn@doi [\mnras] {10.1093/mnras/staa3760}, \href
  {https://ui.adsabs.harvard.edu/abs/2021MNRAS.501.1568F} {501, 1568}

\bibitem[\protect\citeauthoryear{{Gail} \& {Sedlmayr}}{{Gail} \&
  {Sedlmayr}}{1999}]{Gail1999}
{Gail} H.~P.,  {Sedlmayr} E.,  1999, \aap, \href
  {https://ui.adsabs.harvard.edu/abs/1999A&A...347..594G} {347, 594}

\bibitem[\protect\citeauthoryear{{Gall}, {Hjorth}  \& {Andersen}}{{Gall}
  et~al.}{2011}]{Gall2011}
{Gall} C.,  {Hjorth} J.,   {Andersen} A.~C.,  2011, \mn@doi [\aapr]
  {10.1007/s00159-011-0043-7}, \href
  {https://ui.adsabs.harvard.edu/abs/2011A&ARv..19...43G} {19, 43}

\bibitem[\protect\citeauthoryear{{Gomez} et~al.,}{{Gomez}
  et~al.}{2012}]{Gomez2012}
{Gomez} H.~L.,  et~al., 2012, \mn@doi [\mnras]
  {10.1111/j.1365-2966.2011.20272.x}, \href
  {https://ui.adsabs.harvard.edu/abs/2012MNRAS.420.3557G} {420, 3557}

\bibitem[\protect\citeauthoryear{{Hennebelle} \& {Inutsuka}}{{Hennebelle} \&
  {Inutsuka}}{2019}]{Hennebelle2019}
{Hennebelle} P.,  {Inutsuka} S.-i.,  2019, \mn@doi [Frontiers in Astronomy and
  Space Sciences] {10.3389/fspas.2019.00005}, \href
  {https://ui.adsabs.harvard.edu/abs/2019FrASS...6....5H} {6, 5}

\bibitem[\protect\citeauthoryear{{Hildebrandt} et~al.,}{{Hildebrandt}
  et~al.}{2010}]{Hildebrandt2010}
{Hildebrandt} H.,  et~al., 2010, \mn@doi [\aap] {10.1051/0004-6361/201014885},
  \href {https://ui.adsabs.harvard.edu/abs/2010A&A...523A..31H} {523, A31}

\bibitem[\protect\citeauthoryear{{Hopkins}}{{Hopkins}}{2012a}]{imf-t}
{Hopkins} P.~F.,  2012a, \mn@doi [\mnras] {10.1111/j.1365-2966.2012.20731.x},
  \href {https://ui.adsabs.harvard.edu/abs/2012MNRAS.423.2037H} {423, 2037}

\bibitem[\protect\citeauthoryear{Hopkins}{Hopkins}{2012b}]{Hopkins2012}
Hopkins P.~F.,  2012b, \mn@doi [Monthly Notices of the Royal Astronomical
  Society] {10.1111/j.1365-2966.2012.20731.x}, 423, 2037

\bibitem[\protect\citeauthoryear{{Ilbert} et~al.,}{{Ilbert}
  et~al.}{2013}]{Ilbert2013}
{Ilbert} O.,  et~al., 2013, \mn@doi [\aap] {10.1051/0004-6361/201321100}, \href
  {https://ui.adsabs.harvard.edu/abs/2013A&A...556A..55I} {556, A55}

\bibitem[\protect\citeauthoryear{{Jeans}}{{Jeans}}{1902a}]{Jeans1902}
{Jeans} J.~H.,  1902a, \mn@doi [Philosophical Transactions of the Royal Society
  of London Series A] {10.1098/rsta.1902.0012}, \href
  {http://adsabs.harvard.edu/abs/1902RSPTA.199....1J} {199, 1}

\bibitem[\protect\citeauthoryear{{Jeans}}{{Jeans}}{1902b}]{Jeans}
{Jeans} J.~H.,  1902b, \mn@doi [Philosophical Transactions of the Royal Society
  of London Series A] {10.1098/rsta.1902.0012}, \href
  {https://ui.adsabs.harvard.edu/abs/1902RSPTA.199....1J} {199, 1}

\bibitem[\protect\citeauthoryear{{Jermyn}, {Steinhardt}  \& {Tout}}{{Jermyn}
  et~al.}{2018}]{Jermyn2018}
{Jermyn} A.~S.,  {Steinhardt} C.~L.,   {Tout} C.~A.,  2018, \mn@doi [\mnras]
  {10.1093/mnras/sty2123}, \href
  {https://ui.adsabs.harvard.edu/abs/2018MNRAS.480.4265J} {480, 4265}

\bibitem[\protect\citeauthoryear{{Kroupa}}{{Kroupa}}{2001}]{Kroupa2001}
{Kroupa} P.,  2001, \mn@doi [\mnras] {10.1046/j.1365-8711.2001.04022.x}, \href
  {https://ui.adsabs.harvard.edu/abs/2001MNRAS.322..231K} {322, 231}

\bibitem[\protect\citeauthoryear{{Labb{\'e}} et~al.,}{{Labb{\'e}}
  et~al.}{2005}]{Labbe2005}
{Labb{\'e}} I.,  et~al., 2005, \mn@doi [\apjl] {10.1086/430700}, \href
  {https://ui.adsabs.harvard.edu/abs/2005ApJ...624L..81L} {624, L81}

\bibitem[\protect\citeauthoryear{{Laigle} et~al.,}{{Laigle}
  et~al.}{2016}]{Laigle2016}
{Laigle} C.,  et~al., 2016, \mn@doi [\apjs] {10.3847/0067-0049/224/2/24}, \href
  {http://adsabs.harvard.edu/abs/2016ApJS..224...24L} {224, 24}

\bibitem[\protect\citeauthoryear{{Le{\'s}niewska} \&
  {Micha{\l}owski}}{{Le{\'s}niewska} \&
  {Micha{\l}owski}}{2019}]{Lesniewska2019}
{Le{\'s}niewska} A.,  {Micha{\l}owski} M.~J.,  2019, \mn@doi [\aap]
  {10.1051/0004-6361/201935149}, \href
  {https://ui.adsabs.harvard.edu/abs/2019A&A...624L..13L} {624, L13}

\bibitem[\protect\citeauthoryear{{Lonsdale}, {Farrah}  \& {Smith}}{{Lonsdale}
  et~al.}{2006}]{Lonsdale2006}
{Lonsdale} C.~J.,  {Farrah} D.,   {Smith} H.~E.,  2006, {Ultraluminous Infrared
  Galaxies}.
Springer Praxis, p.~285, \mn@doi{10.1007/3-540-30313-8\_9}

\bibitem[\protect\citeauthoryear{{Low} \& {Lynden-Bell}}{{Low} \&
  {Lynden-Bell}}{1976}]{LyndenBell1976}
{Low} C.,  {Lynden-Bell} D.,  1976, \mn@doi [\mnras] {10.1093/mnras/176.2.367},
  \href {http://adsabs.harvard.edu/abs/1976MNRAS.176..367L} {176, 367}

\bibitem[\protect\citeauthoryear{{Magdis} et~al.,}{{Magdis}
  et~al.}{2017}]{Magdis2017}
{Magdis} G.~E.,  et~al., 2017, \mn@doi [\aap] {10.1051/0004-6361/201731037},
  \href {https://ui.adsabs.harvard.edu/abs/2017A&A...603A..93M} {603, A93}

\bibitem[\protect\citeauthoryear{{Magnelli} et~al.,}{{Magnelli}
  et~al.}{2014}]{Magnelli2014}
{Magnelli} B.,  et~al., 2014, \mn@doi [\aap] {10.1051/0004-6361/201322217},
  \href {http://adsabs.harvard.edu/abs/2014A%26A...561A..86M} {561, A86}

\bibitem[\protect\citeauthoryear{{Matsuura} et~al.,}{{Matsuura}
  et~al.}{2011}]{Matsuura2011}
{Matsuura} M.,  et~al., 2011, \mn@doi [Science] {10.1126/science.1205983},
  \href {https://ui.adsabs.harvard.edu/abs/2011Sci...333.1258M} {333, 1258}

\bibitem[\protect\citeauthoryear{{Mowla}, {van der Wel}, {van Dokkum}  \&
  {Miller}}{{Mowla} et~al.}{2019}]{Mowla2019}
{Mowla} L.,  {van der Wel} A.,  {van Dokkum} P.,   {Miller} T.~B.,  2019,
  \mn@doi [\apjl] {10.3847/2041-8213/ab0379}, \href
  {https://ui.adsabs.harvard.edu/abs/2019ApJ...872L..13M} {872, L13}

\bibitem[\protect\citeauthoryear{{Noeske} et~al.,}{{Noeske}
  et~al.}{2007}]{Noeske2007}
{Noeske} K.~G.,  et~al., 2007, \mn@doi [\apjl] {10.1086/517926}, \href
  {https://ui.adsabs.harvard.edu/abs/2007ApJ...660L..43N} {660, L43}

\bibitem[\protect\citeauthoryear{Papadopoulos}{Papadopoulos}{2010}]{Papadopoulos2010}
Papadopoulos P.~P.,  2010, ApJ, 720, 226

\bibitem[\protect\citeauthoryear{{Peng} et~al.,}{{Peng}
  et~al.}{2010}]{Peng2010}
{Peng} Y.-j.,  et~al., 2010, \mn@doi [\apj] {10.1088/0004-637X/721/1/193},
  \href {http://adsabs.harvard.edu/abs/2010ApJ...721..193P} {721, 193}

\bibitem[\protect\citeauthoryear{{Planck Collaboration} et~al.,}{{Planck
  Collaboration} et~al.}{2020}]{Planck2018}
{Planck Collaboration} et~al., 2020, \mn@doi [\aap]
  {10.1051/0004-6361/201833910}, \href
  {https://ui.adsabs.harvard.edu/abs/2020A&A...641A...6P} {641, A6}

\bibitem[\protect\citeauthoryear{{Raskutti}, {Ostriker}  \&
  {Skinner}}{{Raskutti} et~al.}{2016}]{Raskutti2016}
{Raskutti} S.,  {Ostriker} E.~C.,   {Skinner} M.~A.,  2016, \mn@doi [\apj]
  {10.3847/0004-637X/829/2/130}, \href
  {https://ui.adsabs.harvard.edu/abs/2016ApJ...829..130R} {829, 130}

\bibitem[\protect\citeauthoryear{{Salpeter}}{{Salpeter}}{1955}]{Salpeter1955}
{Salpeter} E.~E.,  1955, \mn@doi [\apj] {10.1086/145971}, \href
  {https://ui.adsabs.harvard.edu/abs/1955ApJ...121..161S} {121, 161}

\bibitem[\protect\citeauthoryear{{Santini} et~al.,}{{Santini}
  et~al.}{2014}]{Santini2014}
{Santini} P.,  et~al., 2014, \mn@doi [\aap] {10.1051/0004-6361/201322835},
  \href {https://ui.adsabs.harvard.edu/abs/2014A&A...562A..30S} {562, A30}

\bibitem[\protect\citeauthoryear{{Schnee}, {Li}, {Goodman}  \&
  {Sargent}}{{Schnee} et~al.}{2008}]{Schnee2008}
{Schnee} S.,  {Li} J.,  {Goodman} A.~A.,   {Sargent} A.~I.,  2008, \mn@doi
  [\apj] {10.1086/590375}, \href
  {https://ui.adsabs.harvard.edu/abs/2008ApJ...684.1228S} {684, 1228}

\bibitem[\protect\citeauthoryear{{Schreiber} et~al.,}{{Schreiber}
  et~al.}{2018}]{Schreiber2018}
{Schreiber} C.,  et~al., 2018, \mn@doi [\aap] {10.1051/0004-6361/201833070},
  \href {https://ui.adsabs.harvard.edu/abs/2018A&A...618A..85S} {618, A85}

\bibitem[\protect\citeauthoryear{{Sloan}}{{Sloan}}{2017}]{Sloan2017}
{Sloan} G.~C.,  2017, \mn@doi [\planss] {10.1016/j.pss.2017.07.017}, \href
  {https://ui.adsabs.harvard.edu/abs/2017P&SS..149...32S} {149, 32}

\bibitem[\protect\citeauthoryear{{Sneppen}, {Steinhardt}, {Hensley}, {Jermyn},
  {Mostafa}  \& {Weaver}}{{Sneppen} et~al.}{2022}]{Sneppen2022}
{Sneppen} A.,  {Steinhardt} C.~L.,  {Hensley} H.,  {Jermyn} A.~S.,  {Mostafa}
  B.,   {Weaver} J.~R.,  2022, arXiv e-prints, \href
  {https://ui.adsabs.harvard.edu/abs/2022arXiv220511536S} {p. arXiv:2205.11536}

\bibitem[\protect\citeauthoryear{{Speagle}, {Steinhardt}, {Capak}  \&
  {Silverman}}{{Speagle} et~al.}{2014}]{Speagle2014}
{Speagle} J.~S.,  {Steinhardt} C.~L.,  {Capak} P.~L.,   {Silverman} J.~D.,
  2014, \mn@doi [\apjs] {10.1088/0067-0049/214/2/15}, \href
  {https://ui.adsabs.harvard.edu/abs/2014ApJS..214...15S} {214, 15}

\bibitem[\protect\citeauthoryear{{Steinhardt} et~al.,}{{Steinhardt}
  et~al.}{2014}]{Steinhardt2014a}
{Steinhardt} C.~L.,  et~al., 2014, \mn@doi [\apjl]
  {10.1088/2041-8205/791/2/L25}, \href
  {http://adsabs.harvard.edu/abs/2014ApJ...791L..25S} {791, L25}

\bibitem[\protect\citeauthoryear{{Steinhardt}, {Jermyn}  \&
  {Lodman}}{{Steinhardt} et~al.}{2020a}]{Steinhardt2020a}
{Steinhardt} C.~L.,  {Jermyn} A.~S.,   {Lodman} J.,  2020a, \mn@doi [\apj]
  {10.3847/1538-4357/ab66b7}, \href
  {https://ui.adsabs.harvard.edu/abs/2020ApJ...890...19S} {890, 19}

\bibitem[\protect\citeauthoryear{{Steinhardt}, {Weaver}, {Maxfield},
  {Davidzon}, {Faisst}, {Masters}, {Schemel}  \& {Toft}}{{Steinhardt}
  et~al.}{2020b}]{Steinhardt2020b}
{Steinhardt} C.~L.,  {Weaver} J.~R.,  {Maxfield} J.,  {Davidzon} I.,  {Faisst}
  A.~L.,  {Masters} D.,  {Schemel} M.,   {Toft} S.,  2020b, \mn@doi [\apj]
  {10.3847/1538-4357/ab76be}, \href
  {https://ui.adsabs.harvard.edu/abs/2020ApJ...891..136S} {891, 136}

\bibitem[\protect\citeauthoryear{{Steinhardt}, {Sneppen}, {Hensley}  \&
  {Mostafa}}{{Steinhardt} et~al.}{2022}]{Steinhardt2022b}
{Steinhardt} C.,  {Sneppen} A.,  {Hensley} H.,   {Mostafa} B.,  2022, in prep.

\bibitem[\protect\citeauthoryear{{Stringer}, {Benson}, {Bundy}, {Ellis}  \&
  {Quetin}}{{Stringer} et~al.}{2009}]{Stringer2009}
{Stringer} M.~J.,  {Benson} A.~J.,  {Bundy} K.,  {Ellis} R.~S.,   {Quetin}
  E.~L.,  2009, \mn@doi [\mnras] {10.1111/j.1365-2966.2008.14186.x}, \href
  {https://ui.adsabs.harvard.edu/abs/2009MNRAS.393.1127S} {393, 1127}

\bibitem[\protect\citeauthoryear{{Wild} et~al.,}{{Wild}
  et~al.}{2014}]{Wild2014}
{Wild} V.,  et~al., 2014, \mn@doi [\mnras] {10.1093/mnras/stu212}, \href
  {https://ui.adsabs.harvard.edu/abs/2014MNRAS.440.1880W} {440, 1880}

\bibitem[\protect\citeauthoryear{{Wild} et~al.,}{{Wild}
  et~al.}{2020}]{Wild2020}
{Wild} V.,  et~al., 2020, \mn@doi [\mnras] {10.1093/mnras/staa674}, \href
  {https://ui.adsabs.harvard.edu/abs/2020MNRAS.494..529W} {494, 529}

\bibitem[\protect\citeauthoryear{{Williams}, {Quadri}, {Franx}, {van Dokkum}
  \& {Labb{\'e}}}{{Williams} et~al.}{2009}]{Williams2009}
{Williams} R.~J.,  {Quadri} R.~F.,  {Franx} M.,  {van Dokkum} P.,   {Labb{\'e}}
  I.,  2009, \mn@doi [\apj] {10.1088/0004-637X/691/2/1879}, \href
  {https://ui.adsabs.harvard.edu/abs/2009ApJ...691.1879W} {691, 1879}

\makeatother
\end{thebibliography}

\label{lastpage}
\end{document}